\documentclass[twocolumn,showpacs,preprintnumbers,amsmath,amssymb]{revtex4}
\usepackage{epsfig}

\begin{document}
\draft

\title{Correlations and scaling in one-dimensional heat conduction}

\author{J. M. Deutsch and Onuttom Narayan}
\affiliation{
Department of Physics, University of California, Santa Cruz, CA 95064.}

\date{\today}

\begin{abstract}
We examine numerically the full spatio-temporal correlation functions
for all hydrodynamic quantities for the random collision model introduced
recently.  The autocorrelation function of the heat current, through the
Kubo formula, gives a thermal conductivity exponent of $1/3$ in agreement
with the analytical prediction and previous numerical work.  Remarkably,
this result depends crucially on the choice of boundary conditions: for
periodic boundary conditions (as opposed to open boundary conditions
with heat baths) the exponent is approximately $1/2$.  This is expected
to be a generic feature of systems with singular transport coefficients.
All primitive hydrodynamic quantities scale with the dynamic critical 
exponent predicted analytically.  
\end{abstract}


\maketitle 

\section{Introduction}
Heat conduction in one dimensional systems is a simple example of the
general problem of singular transport coefficients\cite{Lieb}. Apart
from this theoretical significance, there is increasing
experimental relevance due to the tremendous advances in nanotube
technology\cite{Grass2,bhatia}, and the importance of understanding
their thermal properties. It appears that the thermal conductivity
$\kappa(L),$ as a function of the system size $L,$ {\it diverges\/}
with increasing $L$ for one-dimensional momentum conserving
systems\cite{prosen,Grass,Casati,ONSR}. Recently\cite{ONSR}, a
hydrodynamic description has been proposed, which obtained the analytic
prediction of
\begin{equation}
\kappa(L)\sim L^{1/3}
\end{equation}
for such momentum conserving systems\cite{Grass,ONSR}. Numerical results
on various systems~\cite{Grass,Casati}, most recently work on Sinai
and Chernov's pencase model\cite{pencase} and the random collision (RC)
model\cite{DeutschNarayan}, confirm this prediction.

Despite the numerical confirmation, there is reason to investigate the
analytical predictions further. The hydrodynamic description makes
detailed predictions about the dynamical behavior of all conserved
quantities in the system: energy, momentum and mass. The heat conductivity
is only a single, and rather indirect, verification of the theoretical
picture.  In addition, there are significant corrections to the scaling
form predicted for $\kappa(L),$ so that it is appropriate to explore
whether they are indeed simply corrections to scaling for small $L,$
or reflect a departure from the hydrodynamic description.  Finally,
the connection between conductivity and correlation functions through
the Kubo formula\cite{Kubo}, invoked in the analytical treatment~\cite{ONSR},
is worth further examination.

In this paper, we study the RC model\cite{DeutschNarayan}, which was
recently introduced for one dimensional transport, and numerically obtain
the full spatiotemporal form of all correlation functions between the
conserved hydrodynamic variables. We find both propagating and dispersive
modes, with a dispersion relation in agreement with the analytical result.

The autocorrelation function of the heat current, which can be obtained
in terms of the primitive hydrodynamic quantities has a spatiotemporal
structure that is not amenable to any simple scaling description
for length and time scales much larger than (about thirty times) the
interparticle spacing and collision time. However, as we will argue,
the existence of larger corrections to scaling for the heat current
is perhaps not surprising. Moreover, the continuity relation connects
the heat current to the primitive hydrodynamic quantities, so these
corrections must vanish on suitably large length and time scales.
In accordance with this expectation, the autocorrelation function of
the total heat current (integrated over the entire system), $C_{JJ}(t),$
is much better behaved. The integral
\begin{equation}
I(t)\equiv{1\over L}\int_0^t dt^\prime C_{JJ}(t^\prime)
\label{IT}
\end{equation} 
scales with $t$ with an exponent close to $1/3,$ although depending on
the model parameters chosen $L$ can be as large as $\sim 10000$ before
the asymptotic scaling behavior is seen.

The autocorrelation function $C_{JJ}(t)$ was measured with two different
boundary conditions: open boundary conditions with heat baths at the
ends, and periodic boundary conditions. In both cases, $I(t)$ looks
the same at short times, and has oscillatory behavior for $t\sim L.$
For open boundary conditions, the oscillations damp out rapidly,
and therefore asymptote on a time scale $\sim L,$ as assumed in the
analytical treatment. Thus $I(t\rightarrow\infty),$ which is connected
to the conductivity through the Kubo formula, scales as $\sim L^{1/3}.$
On the other hand, with periodic boundary conditions, $I(t)$ continues
to grow after the oscillations set in, and $I(\infty)\sim L^{1/2}.$
This dependence on boundary conditions, although highly unusual, is
consistent with the analytical picture: the dynamical exponent of $3/2$
would normally convert a $t^{1/3}$ scaling to $L^{1/2},$ but with open
boundary conditions the cutoff to the time integral is set by the modes
propagating to the heat baths at the ends, i.e. at $t$ of $O(L).$

\section{Earlier results}
A large number of systems with dimension $d=1$
have been analytically\cite{Exact,Dhar,LLP} or
numerically\cite{Grass2,LLP,Numeric,Grass,Casati} found to have a singular
heat conductivity, $\kappa(L)\sim L^\alpha$ with $\alpha > 0.$ Various
values of $\alpha$ have been obtained for different models. Positive
values of $\alpha$ are seen either when a model is integrable, when
the value of $\alpha$ depends on the details of the model, or when the
interparticle interactions in the model are momentum conserving.

For any non-integrable system, it is possible to construct a hydrodynamic
description, where it is assumed that local thermal equilibrium is
reached\cite{forster,ONSR}. Since long range equilibrium is precluded
in $d=1$, the hydrodynamic description is only in terms of conserved
quantities: energy, number and (when momentum is conserved) momentum
densities. This is the standard hydrodynamic treatment of a normal
fluid. There are two propagating sound modes, and a diffusive mode
for heat transport. The addition of thermal noise to the hydrodynamic
equations leads to singular corrections when $d<2.$ The spreading of a
sound pulse is now superdiffusive: the width $l$ is related to time $t$
as $t\sim l^z,$ with a dynamic exponent $z$ exactly equal to 3/2 for
$d=1.$ The heat diffusion mode is no longer diffusive: contributions
to the heat current that are propagating and bilinear in the primitive
hydrodynamic densities, that are inconsequential for $d > 2,$ dominate
(for long wavelength and low frequency) for $d=1.$ The autocorrelation
function of the heat current decays as $C_{JJ}(t)\sim t^{-2/3}.$ If one
integrates this function, with open boundary conditions and heat baths,
the propagating nature of the modes cuts off the integral $I(t)$ at $t$
of $O(L).$ Using the Kubo formula, one obtains $\kappa(L)\sim L^\alpha$
with $\alpha = 1/3.$

\section{The Model}
There have been numerical studies of heat conduction in various one
dimensional models\cite{Grass2,LLP,Numeric,Grass,Casati}. One of the
simplest such models is that of a chain of point particles undergoing
one dimensional elastic collisions, with heat baths at the two ends of
the chain. The drawback of this model is that it is not 
chaotic\cite{Grass}; in fact, when the particles in the chain are all 
of equal mass, the model is integrable. As explained in the
previous section, the hydrodynamic description that yields a universal
value of $\alpha = 1/3$ is only valid if the system reaches local thermal
equilibrium. Even when the masses of the particles in the chain alternate,
$\alpha$ is found to converge very slowly towards its asymptotic value,
requiring system sizes of $\sim 16,000$ or larger.

To alleviate this problem, we recently introduced the random collision
(RC) model\cite{DeutschNarayan}. This can be viewed as the limit of
rough particles confined to a narrow tube, with the width of the tube
sufficiently small that particles cannot pass each other, in the limit
that the size of the particles (and the width of the tube) approach
zero. In this limit, the particles move along a one dimensional line,
but possess longitudinal as well as transverse momentum, both of
which contribute to the particle energy. Interparticle collisions are
elastic. Owing to the roughness of the particles, particles emerge from
a collision with random momenta (respecting energy momentum conservation
and detailed balance).  The transverse degree of freedom effectively
serves as a random source/sink of longitudinal kinetic energy in any
collision, while strictly maintaining conservation of total energy. The
extra randomization introduced by the transverse momentum is expected
to equilibrate the system much more effectively. Numerically, the
conductivity $\kappa(L)$ is found to scale much better, with $L^\alpha$
scaling observed for $L\sim 100$ to 1000. However, depending on the
particle masses in the model, there is a slight variation in the measured
$\alpha,$ with $\alpha = 0.28$ when all the particles have the same mass,
and $\alpha = 0.33$ when the particle masses alternate with a mass ratio
of 2.62. This indicates that corrections to the asymptotic scaling form,
although much reduced, have not been completely eliminated.

\section{Spatiotemporal correlations}
In this section, we present numerical results for the spatiotemporal 
dependence of the correlation functions of the primitive hydrodynamic
densities: energy, momentum and number. The motivation for carrying 
out this study has been discussed at the beginning of this paper.

\begin{figure}
\centerline{\epsfxsize=\columnwidth \epsfbox{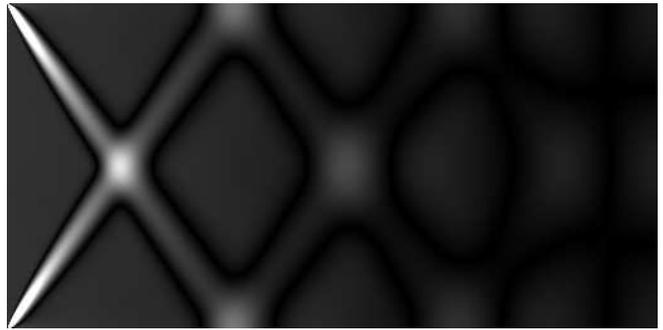}} 
\caption{Grayscale plot of the autocorrelation function of the momentum
density,  $C_{vv}(x, t).$ The vertical and horizontal directions are
$x$ and $t$ respectively. Periodic boundary conditions are used for the
$x$ direction. The length of the system is $L=512.$ The bright regions
correspond to intensity peaks, and clearly show the propagating sound
modes.}
\label{grayscale} 
\end{figure}
\begin{figure}
\centerline{\epsfxsize=\columnwidth \epsfbox{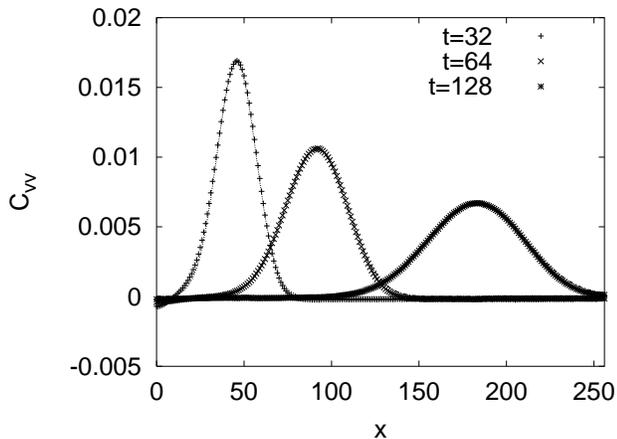}}
\caption{Autocorrelation function of the momentum density,
$C_{vv}(x, t),$ as a function of $x$ for $t=32,$ 64 and 128. Periodic
boundary conditions are used for $x,$ and the length of the system is 
$L=512.$ Since $C_{vv}$ is symmetric in $x,$ only $x>0$ is shown. 
As is shown in the figure, the data fits a Gaussian form excellently, 
propagating outwards at constant speed and broadening as $\sim t^{2/3}.$}
\label{mom-mom} 
\end{figure}

Figure~\ref{grayscale} shows  the autocorrelation function of the
momentum density, $C_{vv}(x, t),$ as a function of position $x$ and
time $t$.  The size of the system is $L=512,$ all particles have unit
mass, and periodic boundary conditions in $x$ are used.  One can clearly
see two pulses propagating in opposite directions at constant speed,
corresponding to the two sound modes. Even when the pulses collide
with each other, they emerge essentially unaffected. The figure also
shows a slight broadening of the pulses as $t$ is increased. To make
this picture quantitative, Figure~\ref{mom-mom} plots $C_{vv}(x, t)$
as a function of $x$ for various values of $t.$
The lines going through the data points are fits to the functional form
$C_{vv}(x, t) = (A/t^{2/3}) \exp[-B(x-ct)^2/t^{4/3}],$ with $A, B$ and $c$
as fitting parameters that are independent of $t$~\cite{background}. The
fitting form works surprisingly well. Although for spatial dimension
$d> 2$ one indeed expects a propagating Gaussian pulse, for $d < 2$
the scaling theory only predicts that the pulse should spread out as
$\sim t^{2/3},$ and does not require a Gaussian form.
\begin{figure}
\centerline{\epsfxsize=\columnwidth\epsfbox{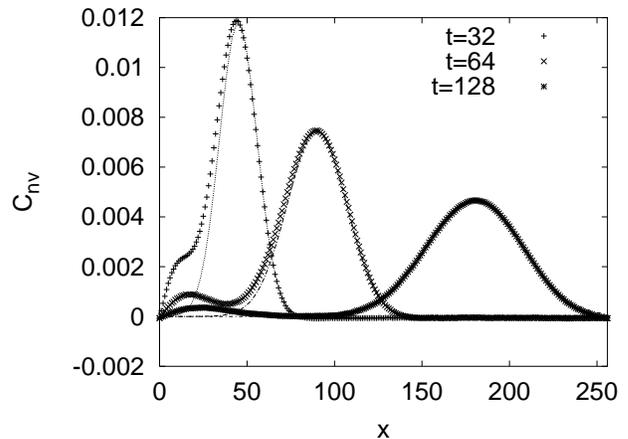}}
\caption{Crosscorrelation function of the mass and momentum density,
$C_{nv}(x, t),$ as a function of $x$ for $t=32,$ 64 and 128. Periodic
boundary conditions are used for $x,$ and the length of the system is 
$L=512.$ Only $x>0$ is shown, since the function is antisymmetric.
The large propagating peak, corresponding to the sound mode, is fitted
to a Gaussian in the figure. The bump near the origin comes from the 
heat mode, and has not been fitted.}
\label{num-mom}
\end{figure}
Similar results are shown in Figure~\ref{num-mom} for the cross
correlation function of the number and momentum densities, $C_{nv}(x, t).$

\begin{figure}
\centerline{\epsfxsize=\columnwidth\epsfbox{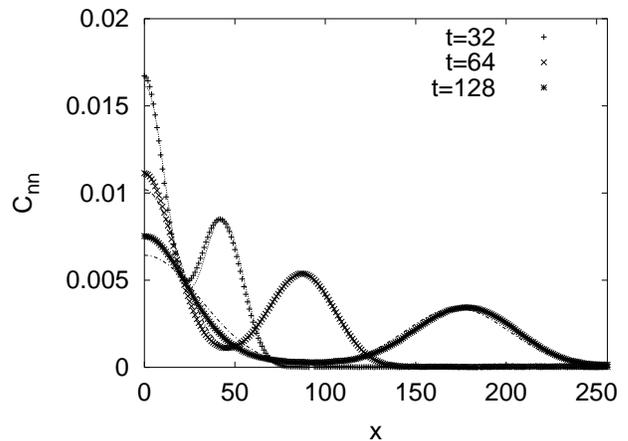}}
\caption{Autocorrelation function of the mass density, $C_{nn}(x, t),$
as a function of $x$ for $t=32,$ 64 and 128. Periodic boundary conditions
are used for $x,$ and the length of the system is $L=512.$ The function
is fitted to the sum of a propagating and diffusing Gaussian, with the
width of both scaling like $\sim t^{2/3}.$.  The second peak does not
fit as well as the first one; as discussed in the text, the corrections
to scaling for this peak are expected to be much stronger.}
\label{num-num}
\end{figure}
Figure~\ref{num-num} shows the autocorrelation function for the number
density $C_{nn}(x, t).$ In addition to the propagating parts, one can
see a peak near the origin. This peak comes from the heat transport
mode, which is diffusive for $d>2.$ The same structure is observed in
the autocorrelation function for the energy density $C_{ee}(x, t).$
\begin{figure}
\centerline{\epsfxsize=\columnwidth\epsfbox{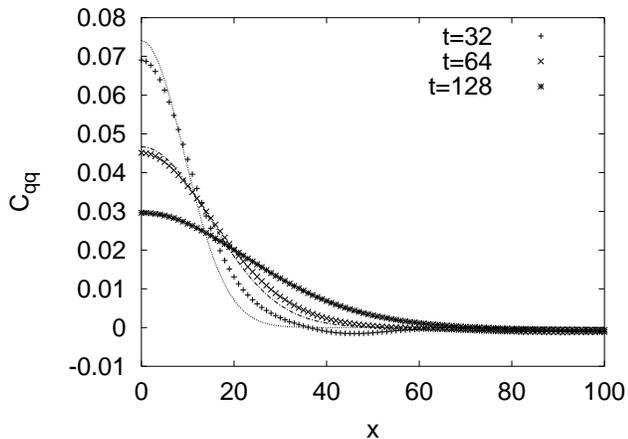}}
\caption{Autocorrelation function for the heat transport mode, Periodic
boundary conditions are used for $x,$ and the length of the system is
$L=512.$ Comparing with $C_{nn}(x, t)$ one can see that the sound peak
has been eliminated. The fit shown in the figure is to a diffusing form:
a Gaussian that is centered at the origin, spreading out as $t^{2/3}.$
The fit is only approximate, as was the case for the peak at the origin
for $C_{nn} in Figure~\ref{num-num}.$}
\label{q-q}
\end{figure}
In order to extract the heat transport mode, suppressing the sound
modes, in Figure~\ref{q-q} we plot $C_{qq}(x, t),$ the autocorrelation
function of 
\begin{equation}
q(x, t) = e(x, t) - (\overline h/\overline n) n(x, t),
\end{equation}
where $\overline h$ and $\overline n$ are the spatially averaged enthalpy
and number density respectively. At long wavelengths and low frequencies,
this corresponds to the heat transport mode~\cite{forster}. The peak
broadens with time with a width $\sim t^{2/3},$ but the fit is not as
good as for the previous plots.

The scaling of $C_{qq}(x, t)$ requires careful consideration for $d <
2.$ The hydrodynamic equation for $q(x, t)$ is
\begin{equation}
\partial_t q + \nabla\cdot
(v \delta q + v a \delta e ) = \kappa_0\nabla^2 T + O(v^2)
\label{heatcond}
\end{equation}
where $\kappa_0$ is the bare heat conductivity, $T$ is the temperature,
$\delta q, \delta e$ are the deviations of $q, e$ from their average
values, and $a = \overline h/\overline e - 1.$ (For the RC model,
$a=1.$) When fluctuations around average values are small, the second
term on the left hand side of Eq.(\ref{heatcond}) is second order
in the fluctuating fields and can be neglected, yielding diffusive
transport for $q.$ However, for $d < 2,$ one can see~\cite{ONSR}
that this advective contribution to the heat current, $v \delta q +
v a \delta e,$ dominates at long wavelengths and low frequencies. Thus
one expects $q$ to propagate, with the propagating behavior becoming
increasingly important at long length and time scales. This propagating
behavior was used in Ref.~\cite{ONSR} to argue for a cutoff of $t\sim L$
for correlations in a system of size $L$ with heat baths at the ends.
We will discuss this issue further in the next section.

%Figure~\ref{q-q-Fourier} shows
%$C_{qq}(k,\omega)$ in the Fourier domain, as a function of $\omega/k$
%for different values of $k.$ A peak at non-zero $\omega/k$ is seen, whose
%location is approximately independent of $k.$ In the next section, we will
%return to this propagating behavior of $q$ when we measure the finite-size
%cutoffs directly for the autocorrelation function of the heat current.

\section{Heat current}
We have seen in the previous section that the conserved hydrodynamic
quantities satisfy the predictions of the scaling theory: two propagating
modes which disperse according to $\delta x\sim t^{2/3},$ and a heat
transport mode. We have also seen that the third mode, unlike for $d >
2,$ is not a simple diffusing mode. We now turn to the autocorrelation
function of the heat current. For momentum conserving systems, it
can be shown on very general grounds~\cite{mclennan} that this is the
correlation function that is related to the thermal conductivity through
the Kubo formula~\cite{Kubo}.

\begin{figure}
\centerline{\epsfxsize=2.5in\epsfbox{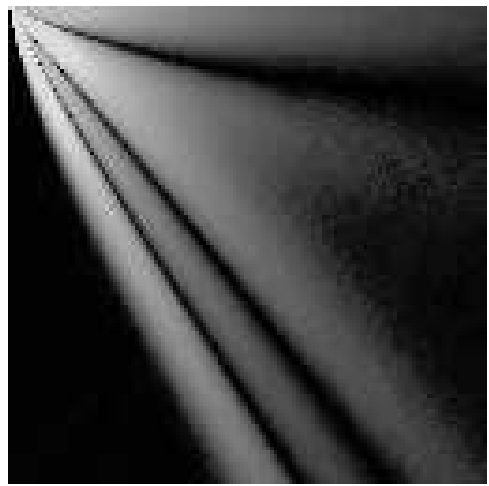}}
\caption{Grayscale plot of the autocorrelation function of the
heat current,  $C_{jj}(x, t).$ The vertical and horizontal directions
are $x$ and $t$ respectively. Periodic boundary conditions are used
for the $x$ direction. The length of the system is $L=512.$ The 
intensity peaks move out linearly from $x=0$ as a function of time.}
\label{jjgray}
\end{figure}
\begin{figure}
\centerline{\epsfxsize=\columnwidth\epsfbox{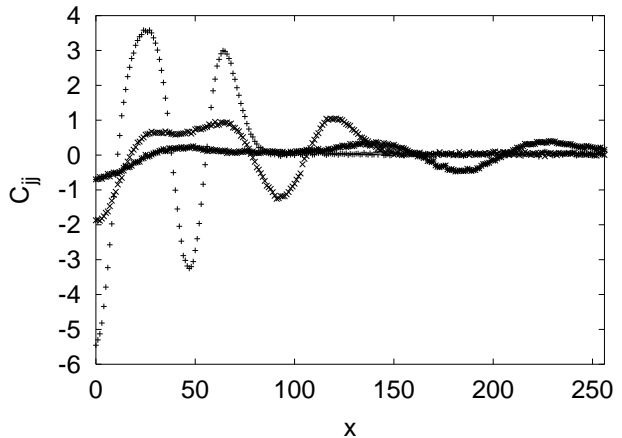}}
\caption{Autocorrelation function of the heat current,
$C_{jj}(x, t),$ as a function of $x$ for $t=32,$ 64 and 128. Periodic
boundary conditions are used for $x,$ and the length of the system is
$L=512.$ The peaks decay, broaden, and shift outwards as a function 
of time. No simple scaling form fits the data.}
\label{currcurr}
\end{figure}

Figure~\ref{jjgray} shows a grayscale plot of the autocorrelation function
of the heat current, $C_{jj}(x, t),$ similar to that for $C_{vv}$ shown in 
the previous section. The current $j(x, t)$ is defined as
\begin{equation}
j(x, t) =\sum_i \delta(x_i - x)v_i [\epsilon_i - (\overline
h/\overline n)]
\label{jmicro}
\end{equation}
where the sum is over particles and $\epsilon_i$
is the energy of the $i$'th particle.  $j(x, t)$ is the current that
corresponds to $q;$ its hydrodynamic form can be obtained from
Eq.(\ref{heatcond}) as
\begin{equation}
j(x, t) = v[\delta q + \delta \epsilon] - \kappa_0\nabla T + O(v^2).
\label{current}
\end{equation}
The propagating nature of the heat current can be seen by following the
intensity peaks in the figure; the range in time over which this can
be seen is not as large as for $C_{vv},$ because $C_{jj}$ decays much
faster. As in the previous section, Figure~\ref{currcurr} shows $C_{jj}$
as a function of $x$ for different values of $t.$ 
Although the peaks shift outwards and broaden with time, unfortunately no
clear scaling of any form can be seen at small $t$.  This result might
seem disturbing, since the values of $t$ in Figure~\ref{currcurr},
though not large, are sufficient to see good scaling in the basic
conserved quantities, as shown in the previous section. However, as seen
in Eq.(\ref{current}), the hydrodynamic form of the current has advective
contributions that are bilinear in the primitive conserved quantities,
which dominate at long length and time scales, and other dissipative
contributions as well.  In a diagrammatic field theoretic expansion,
one expects relatively larger contributions from short wavelength modes,
aggravated by the fact that although the first term on the right hand
side of Eq.(\ref{current}) dominates the asymptotic scaling behavior
of $j(x,t),$ it is second order in fluctuations and will therefore have
large corrections to scaling~\cite{qomega}.  (The correlation function
$C_{jj}(x, t)$ decays so quickly with $t$ that it is impossible to
measure it for much larger $t$ than shown in the figure.)

Since we have argued that the lack of scaling seen for $C_{jj}(x, t)$
is a result of strong subleading corrections to its asymptotic behavior,
it is reasonable to examine its spatial integral, i.e. $C_{jj}$ at zero
wavevector. An added motivation is that, if $J(t) = \int dx j(x, t),$
the Kubo formula obtains the conductivity as
\begin{equation}
\kappa = {\beta^2\over L}\int_0^t dt^\prime C_{JJ}(t^\prime)
\end{equation}
where $\beta$ is the inverse temperature.  From the definition of
Eq.(\ref{IT}), the right hand side of this equation is $\beta^2 I(t).$
We first show the results for $I(t)$ for somewhat different conditions
than those used so far in this paper.

\begin{figure}
\centerline{\epsfxsize=\columnwidth\epsfbox{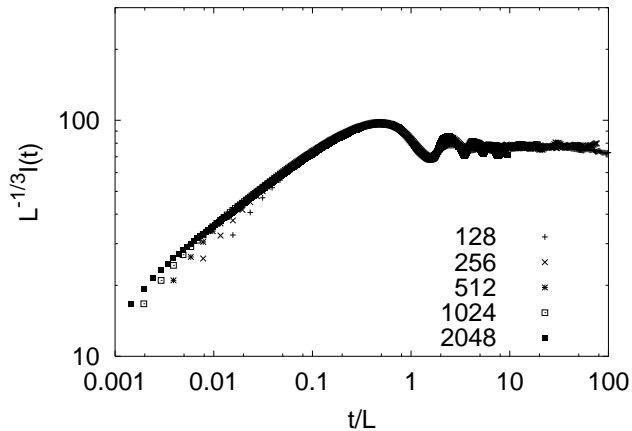}}
\caption{Log log plot of  $I(t)/L^{1/3}$ versus $t/L$ for systems of
different lengths, where $I(t)$ is defined in Eq/(\ref{IT}). The particle
masses alternate between 1 and 2.62. Open boundary 
conditions with heat baths at the two ends are used. The data show a 
good scaling collapse, indicating that $I(\infty)\sim L^{1/3}.$}
\label{kubo-bath-m1m2}
\end{figure}
Figure~\ref{kubo-bath-m1m2} is
a log-log plot of $I(t)/L^{1/3}$ as a function of $t/L$ for different
values of $L.$ Unlike the data shown before this, the system does not
have periodic boundary conditions, but is terminated by heat baths (both
at unit temperature) at both ends.  Also, the masses of the particles
are not all equal, but alternate between 1 and 2.62~\cite{caveat}. The
figure shows a power law rise for $I(t)$ with exponent $t^{1/3},$
which saturates at $t\sim L.$ This is in accordance with the analytical
prediction~\cite{ONSR}. There are a few damped oscillations at integer
multiples of $t\sim L,$ presumably corresponding to successive reflections
from the ends of the system.

\begin{figure}
\centerline{\epsfxsize=\columnwidth\epsfbox{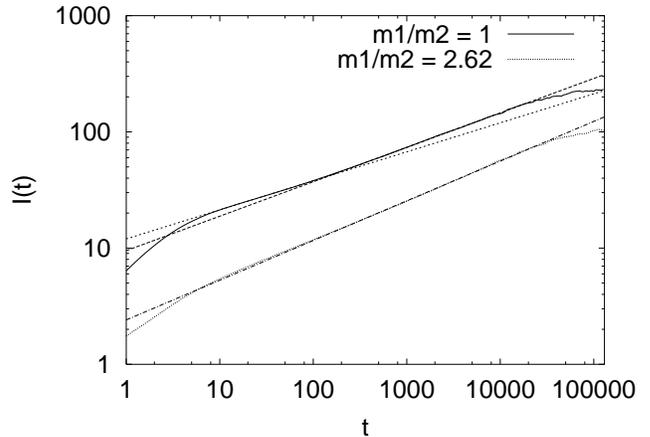}}
\caption{Log log plot of  $I(t)$ versus $t$ for $L=16384,$
where $I(t)$ is defined in Eq/(\ref{IT}). 
All particles have unit mass. The slope shows an upward drift, 
from 0.24 (from
$t=10$ to $t=100$) to 0.29 (from $t=400$ to $t=20000$),
consistent with an eventual asymptotic slope of 1/3. The figure also
shows a similar plot for a chain whose masses alternate between 1 and 2.62,
shifted downwards for clarity. A fit to a slope of 0.33 is shown; no
upward (or downward) drift in the slope is seen.
Periodic boundary conditions are used; over the range fitted, 
the results are the same with open boundary conditions.}
\label{kubo-bath-eq}
\end{figure}
When all the particles in the chain are taken to have equal mass,
there is a slight discrepancy from the analytical prediction.
Figure~\ref{kubo-bath-eq} shows a log-log plot of $I(t)$ for this case.
The power law growth of $I(t)$ has an exponent close to 0.25, but
the saturation at $t\sim L$ is still very clear. Taken at face value,
this exponent would imply a conductivity exponent of $\alpha = 0.25,$
in contradiction to the analytical prediction. However there are two
reasons why this must be regarded simply as lack of complete convergence
to the asymptotic form. Firstly, as seen in this paper, the dynamics of
the primitive hydrodynamic variables {\it do\/} agree with the analytical
predictions; a fit with $\delta x\sim t^{0.75}$ does not work. Secondly,
and more directly, simulations on much larger systems, with $L = 16000,$
show an upward drift of the slope of $\ln I(t)$ versus $\ln t,$ towards
$1/3.$ No such drift of the slope is seen for the unequal mass case
shown in Figure~\ref{kubo-bath-eq}. Note that there are also deviations
from the asymptotic value of $\alpha$ for the equal mass chain when the
conductivity is measured directly~\cite{DeutschNarayan}, which was in
fact the motivation for the detailed spatiotemporal measurements reported
in this paper.

The most unusual result for the the heat current autocorrelation function
is obtained for $I(t)$ for the case of periodic boundary conditions.
\begin{figure}
\centerline{\epsfxsize=\columnwidth\epsfbox{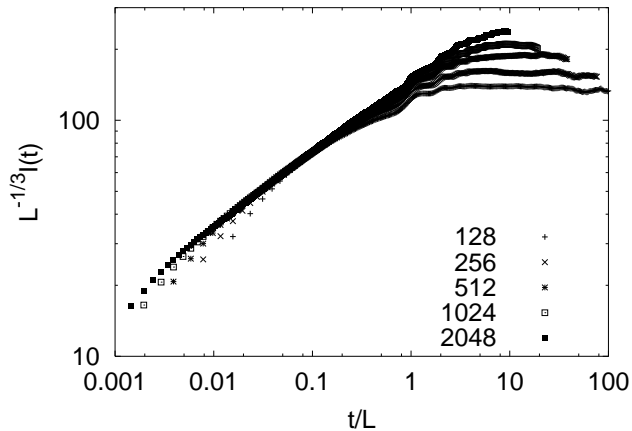}}
\caption{Log log plot of  $I(t)/L^{1/3}$ versus $t/L$ for systems of
different lengths, where $I(t)$ is defined in Eq/(\ref{IT}). The particle
masses alternate between 1 and 2.62. Periodic
boundary conditions are used. For small $t,$ $I(t)$ is the same as 
for open boundary conditions (shown in Figure~\ref{kubo-bath-m1m2}). 
However, beyond the oscillatory regime at $t\sim O(L),$ $I(t)$ keeps
rising, and $I(\infty)$ does not scale as $\sim L^{1/3}.$}
\label{kubo-periodic-m1m2}
\end{figure}
Figure~\ref{kubo-periodic-m1m2} is a log-log plot of $I(t)/L^{1/3}$ as
a function of $t/L.$ As was the case with heat baths at the ends, $I(t)$
grows initially with the form $\sim t^{1/3},$ has damped oscillations at
$t\sim L,$ and saturates for large $t.$ However, {\it unlike\/} the case
with heat baths, $I(t)$ continues to grow even after the oscillations
set in, and the eventual asymptotic value scales approximately like
$L^{1/2}.$ In hindsight, this result is reasonable: without the
heat baths, the dynamics are not cut off at the ends of the system,
and the dynamic scaling exponent of $z = 3/2$ saturates the $t^{1/3}$
growth at $\sim L^{1/2}.$ Alternatively, the saturation of $I(t)$ is
set by the time a pulse takes to spread across the system rather than
propagate across it. It is not clear exactly why the spreading cuts off
the growth of $I(t),$ though from a scaling viewpoint one could argue
that in the absence of the propagation timescale, the only possibility
is $L\sim t^{2/3}.$
\begin{figure}
\centerline{\epsfxsize=2.5in\epsfbox{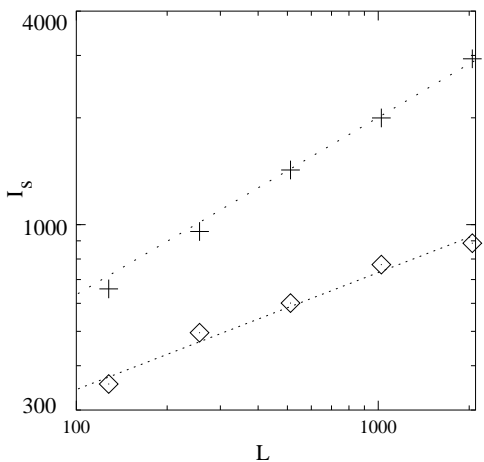}}
\caption{Log log plot of  $I(\infty)$ versus $L$ for open and 
periodic boundary conditions. The straight lines show fits to 
$\sim L^{1/3}$ and $\sim L^{1/2}$ respectively.}
\label{slope}
\end{figure}

\section{Discussion}
We have seen that the dynamics of the primitive hydrodynamic quantities,
mass, momentum and energy density, fit a scaling description with
the predicted dynamic exponent. The scaling of the heat current
autocorrelation function is more problematic. However, we have argued
that this is to be expected, because the asymptotically dominant part of
the current operator is bilinear in the primitive quantities and has a
small bare value in the sense of the renormalization group compared to
correction terms. The same slow convergence to asymptotic scaling that
afflicts the heat current autocorrelation function has been noticed for
the thermal conductivity and remarked on by various authors. However, the
results for the primitive hydrodynamic quantities make it clear that this
{\it is\/} indeed simply a case of slow convergence to the asymptotic
limit rather than a discrepancy with the analytical prediction. Most
remarkably, the large time limit of the spatiotemporal integral $I(t)$
of the heat current autocorrelation function scales with $L$ as $\sim
L^{1/3}$ for a system terminated with heat baths, and as $\sim L^{1/2}$
for periodic boundary conditions. This implies that in applying the
Kubo formula to systems with singular transport coefficients one must
be careful about the boundary conditions being used.

\section{Acknowledgements}
It is a pleasure to acknowledge useful discussions with Abhishek Dhar and 
Sriram Ramaswamy.


\begin{references}
\bibitem{Lieb}
Z. Rieder, J.L. Lebowitz and E. Lieb, J. Math. Phys. {\bf 8}, 1073 (1967).

\bibitem{Grass2} P. Grassberger and L. Yang, cond-mat/0204247

\bibitem{bhatia} S.K. Bhatia and D. Nicholson, Phys. Rev. Lett. {\bf 90} 016105 (2003).

\bibitem{prosen} T. Prosen and D.K. Campbell, Phys. Rev. Lett. {\bf 84},
2857 (2000).

\bibitem{Grass} P. Grassberger, W. Nadler and L. Yang, Phys. Rev. Lett.
{\bf 89}, 180601 (2002).


\bibitem{Casati} G. Casati and T. Prosen, cond-mat/0203331


\bibitem{ONSR} O. Narayan and S. Ramaswamy, Phys. Rev. Lett. {\bf 89}, 200601
(2002).


\bibitem{pencase} Ya. G. Sinai and N.I. Chernov, Russian Math. Surveys
{\bf (3) 42}, 181 (1977).


\bibitem{DeutschNarayan} J.M. Deutsch and O. Narayan cond-mat/0301181;
Phys. Rev. E (Rapid Communications), in press. 

\bibitem{Kubo} R. Kubo, J. Phys. Soc. Japan {\bf 12}, 570 (1957).


\bibitem{Exact}
A. Casher and J.L. Lebowitz, J. Math. Phys. {\bf 12}, 1701
(1971); R.J. Rubin and W.L. Greer, J. Math. Phys. {\bf 12}, 1686 (1971);
A.J. O'Connor and J.L. Lebowitz, J. Math. Phys. {\bf 15}, 692 (1974);
H. Spohn and J.L. Lebowitz, Commun. Math. Phys. {\bf 54}, 97 (1977);
H. Matsuda and K. Ishii, Prog. Theor. Phys. Suppl. {\bf 45}, 56 (1970).

\bibitem{Dhar} A. Dhar, Phys. Rev. Lett. {\bf 86}, 5882 (2001).

\bibitem{LLP} S. Lepri, R. Livi and A. Politi, Europhys. Lett. {\bf 43},
271 (1998).



\bibitem{Numeric} S. Lepri, R. Livi and A. Politi, Phys. Rev. Lett. {\bf
78}, 1896 (1997); A.V. Savin, G.P. Tsironis, A.V. Zolotaryuk,
Phys. Rev. Lett. {\bf 88}, 154301 (2002); T. Hatano, Phys. Rev. E {\bf
59}, R1 (1999); A. Dhar Phys. Rev. Lett. {\bf 86}, 3554 (2001).

\bibitem{forster} "Hydrodynamic Fluctuations, Broken Symmetry, and Correlation 
Functions" D. Forster, Perseus Publishing (1994).


\bibitem{background} A small background has to be subtracted from
$C_{pp},$ because for any given system the center of mass velocity is
conserved. Thus even though in the full canonical ensemble $C_{pp}(x,
t) = 0$ for large $x,$ for zero center of mass velocity (for which the
data is shown) there is a background value that ensures that $\int dx
C_{pp}(x, t) = 0.$

\bibitem{mclennan} J.A. McLennan, {\it Non-equilibrium statistical
mechanics\/}, (Prentice Hall, Engelwood Cliffs NJ 1989); see also
F. Bonetto J.L. Lebowitz and L. Rey-Bellet, math-ph/0002052.

\bibitem{qomega} One might alternatively try to obtain $C_{jj}$ through
the conservation law: $C_{jj}(k, \omega) = (\omega^2/k^2) C_{qq}(k, 
\omega).$ However this is a singular transformation, and relatively small
corrections to scaling in $C_{qq}$ can be accentuated in $C_{jj}.$ We
have verified that our numerical result for $C_{jj}(k, \omega)$ is 
consistent with $(\omega^2/k^2)C_{qq}(k, \omega),$  but this does not
lead to a good scaling form for $C_{jj}.$

\bibitem{caveat} In this case, one has to subtract $(\overline h/\overline m)
g(x, t)$ from the energy current, where $\overline m$ is the average mass 
density and $g(x, t)$ is the momentum density. Thus the second term on 
the right hand side of Eq.(\ref{jmicro}) is changed to 
$(\overline h/\overline m)\sum_i \delta(x - x_i) m_i v_i.$
\end{references}
\end{document}